\documentclass[review]{siamart1116}

\usepackage{lipsum}
\usepackage{amsfonts}
\usepackage{graphicx}
\usepackage{epstopdf}
\usepackage{algorithmic}
\ifpdf
  \DeclareGraphicsExtensions{.eps,.pdf,.png,.jpg}
\else
  \DeclareGraphicsExtensions{.eps}
\fi

\numberwithin{theorem}{section}

\newcommand{\TheTitle}{Nematic director fields and topographies of shells of revolution} 
\newcommand{\TheAuthors}{Mark Warner, and Cyrus Mostajeran}

\newcommand{\ShortTitle}{} 

\headers{\ShortTitle}{\TheAuthors}

\title{{\TheTitle}}

\author{
  Mark Warner\thanks{Cavendish Laboratory, University of Cambridge, 19 JJ Thomson Avenue, Cambridge CB3 0HE, United Kingdom (\email{mw141@cam.ac.uk})}
  \and
    Cyrus Mostajeran\thanks{Department of Engineering, University of Cambridge, United Kingdom (\email{csm54@cam.ac.uk})}
}

\usepackage{amsopn}

\ifpdf
\hypersetup{
  pdftitle={\TheTitle},
  pdfauthor={\TheAuthors}
}
\fi

\def\half{{\textstyle \frac{1}{2}}}
\def\third{{\textstyle \frac{1}{3}}}
\def\quarter{{\textstyle \frac{1}{4}}}
\def\eighth{{\textstyle \frac{1}{8}}}
\def\threehalf{{\textstyle \frac{3}{2}}}
\def\n{\boldsymbol{n}}
\def\r{\boldsymbol{r}}
\def\t{\boldsymbol{t}}
\def\d{{\rm d }}
\def\e{{\rm e }}
\def\N{{\boldsymbol{N} }}
\def\w{\boldsymbol{w}}
\def\F{\boldsymbol{F}}
\def\a{\boldsymbol{a}}

\def\s#1{_{\textrm{#1}}}

\def\be{\begin{equation}}
\def\ee{\end{equation}}
\def\bea{\begin{eqnarray}}
\def\eea{\end{eqnarray}}

\def\det#1{{\rm Det}( #1 )} 

\begin{document}

\begin{nolinenumbers}

\maketitle

\begin{abstract}
We solve the forward and inverse problems associated with the transformation of flat sheets  to surfaces of revolution with non-trivial topography, including Gaussian curvature, without a stretch elastic cost.  We deal with systems slender enough to have a small bend energy cost. Shape change is induced by light or heat causing contraction along a non-uniform director field in the plane of an initially flat nematic sheet.
The forward problem is, given a director distribution, what shape is induced?  Along the way, we determine the Gaussian curvature and the evolution with induced mechanical deformation of the director field and of material curves in the surface (proto-radii) that will become radii in the final surface. The inverse problem is, given a target shape, what director field does one need to specify? Analytic examples of director fields are fully calculated that will, for specific deformations, yield catenoids and paraboloids of revolution. The general prescription is given in terms of an integral equation.

\end{abstract}  


\section{Introduction}

We explore the creation of intrinsically curved shells from flat sheets, completing the connection between the specification of planar mechanical response to the final shapes obtaining. Mechanical response due to heat, light or solvent uptake that spatially varies in-plane means that the metric, specifying intrinsic lengths in a sheet, also varies. Such metric variation means that lengths in neighbouring elements in a plane that become inconsistent with each other can only be resolved by a topography change to that of a Gaussian-curved state. The simplest example is discussed in \cite{modes2010disclinations,modes2011gaussian}, in the context of circular symmetry that interests us here, that is of length changes along and perpendicular to a nematic director, $\boldsymbol{n}$ of a liquid crystalline solid, but where $\n$ forms concentric circles: Contraction along the preferred (circumferential) direction and (radial) elongation perpendicular leads to the ratio of circumference and radius deviating from $2\pi$, necessitating the formation of cones (with localised Gaussian curvature, GC, at their tips) in order to avoid in-plane stretches.

The connection between metric change and GC is straightforward in principle, though perhaps in practice arduous to calculate, see \cite{aharoni2014geometry,Mostajeran2015,Mostajeran2016} for examples in nematic solids. In particular see equations (2.2) and (3.6) of \cite{Mostajeran2016} for the prescription in the current context for connecting this metric variation with GC.  However connecting GC with the topography of a final state without stretch (but with, cheaper, bend) is not always possible, is not necessarily 1:1, and is generally very difficult. The final topography is the most interesting aspect of local length changes, and we give an explicit connection for the important case of circularly-symmetric distributions of GC leading to cylindrically-symmetric topographies.  We also thereby show the inverse -- of how to connect a given topography back to its GC and thence to a particular form of metric variation, and hence ultimately to the director distribution required to create the desired Gaussian-curved shape. This connection, forwards and backwards, is also addressed for cylindrical symmetry in some generality by Aharoni {\it et al} \cite{aharoni2014geometry} who, as in \cite{Mostajeran2015,Mostajeran2016}, gave examples of constant curvature surfaces (spherical spindles). We give a straightforward method and illustrate it with some examples beyond constant curvature.

If the extent of solvent absorption is modulated in-plane, even where the swelling is isotropic, GC can arise \cite{klein2007shaping,kim2012designing}. But when the length changes at each point can differ as well in each direction, the topographical response can be much richer.  Such examples arise, irreversibly, in the growth of some plant leaves where the anisotropy of growth direction can vary spatially to give wrinkling \cite{dervaux2008morphogenesis}.

We mentioned simple circular patterns above giving cones.  Logarithmic spirals (where the director has a fixed direction to a radius) also give cones, and have been experimentally investigated by Broer {\it et al} and by White {\it et al} \cite{de2012engineering,ware2015programmable}. Other suitable spirals also give surfaces of constant, but now non-zero curvature (spheres and spherical spindles) \cite{aharoni2014geometry,Mostajeran2015}, and these too are found experimentally \cite{Mostajeran2016}. Such systems have also been explored numerically by R. Selinger {\it et al} \cite{R-Selinger2016} for liquid crystal elastomers where a range of curvatures from complex director distributions lead to shells with curvature.

The imposed deformation of conventional flat sheets to intrinsically curved surfaces without tearing or wrinkling is impossible; it is the map-maker's problem. These non-isometric problems, including for origami, are intriguing in their own right and are reviewed in this context \cite{modes2016review} and have recently indeed been termed ``non-isometric origamis" \cite{Plucinsky_2016}. There is a more practical interest: Impeding the evolution of such non-isometric topographies gives effective stretches and compressions away from induced equilibrium values, leading to strong forces (as opposed to those arising when impeding bend), and thus highly effective actuation. An example is the resting of a load on an array of would-be cones or square pyramids. To avoid in-plane stretches, the load must be lifted on the tips of the shapes evolving below it \cite{ware2015voxelated}, such a lifter being able to lift loads greatly exceeding ($\times 100$ or more) its own weight. Although these structures are slender, and are acting in a pushing mode, they are not susceptible to Euler buckling instabilities due to their pushing through their intrinsic curvture.  Thus actuations large compared with the material thickness are possible, in distinction to more conventional pushers that need to be stout. Equally, a deforming shell anchored at its boundary (see \cite{Mostajeran2016} for a scheme for achieving fixed boundaries, and below for induced paraboloids) will raise a fluid underneath it by suction -- a pump -- or strongly block a channel on deforming into it -- the first valve working by push has been developed by S\'anchez {\it et al} \cite{sanchez2011valve}, albeit of the stout type. In general, we envisage action in these curvature-changing systems to be uniquely functional at the micro-level since there the concept of ``The Material Is the Machine"
 \cite{Bhattacharya2005machine} is very powerful.

\section{Connecting director patterns to topography}
We consider nematic elastomeric or glassy solids with an anisotropy direction $\boldsymbol{n}$  along which there may be elongation or contraction by a factor of $\lambda > 1$ or $\lambda < 1$, respectively. The deformation gradient may be $1/4 < \lambda <4$ for elastomers and $0.9 < \lambda < 1.1$ for glasses,  can be induced by heat or light, and is reversible \cite{warner2003liquid,van2007glassy}. The two directions perpendicular to $\n$ suffer $\lambda^{-\nu}$ where $\nu$ is the thermal or optical equivalent of the Poisson ratio, that is gives the sympathetic transverse mechanical response to a $\lambda$ induced along $\n$; it takes the value $\nu = 1/2$ for elastomers and $\nu \sim 0.8 \dashleftarrow\dashrightarrow 2.0$ for glasses. The thermo-optical deformation gradient tensor is accordingly
\begin{equation}
\F=(\lambda-\lambda^{-\nu})\boldsymbol{n}\otimes\boldsymbol{n}+\lambda^{-\nu}\,\mathrm{Id}_3,
\end{equation}
where $\mathrm{Id}_3$ denotes the identity operator on $\mathbb{R}^3$, and where $\F$ gives the heated or irradiated relaxed state without stresses. We either (i) take an in-plane pattern for $\n(\r)$ to obtain a GC and topography, or (ii) address the inverse problem of a target topography and inquire as to what imprinted in-plane variation of $\n(\r)$ is needed to generate it.

The 2D metric tensor of the deformed sheet
upon stimulation is $\a=\F^T\cdot \F$, that is:
\begin{equation}
\a=(\lambda^2-\lambda^{-2\nu})\boldsymbol{n}\otimes\boldsymbol{n}+\lambda^{-2\nu}\,\mathrm{Id}_2.
\end{equation}
With respect to local coordinates $(x^i)_{i=1,2}$, the metric tensor can be symbolically represented by the squared length element $ds^2=a_{ij}dx^idx^j$, where the metric components take the form
\begin{equation}
a_{ij}=\left(\lambda^2-\lambda^{-2\nu}\right)n_{i}
n_{j}+\lambda^{-2\nu}\delta_{ij}.
\end{equation}
It describes a new system, if it is thin enough to relax stretches in preference to much cheaper bends. The variation of $\a$ gives the connections $\Gamma^{k}_{ij}$. The variation of the connections, and their products with themselves, give $K(\r)$, the GC; see \cite{Mostajeran2016} for these results in this context.

Dealing with circular symmetry, we resort to circular coordinates $(r,\theta)$. The angle $\alpha$ between the director and the radius is only a function of $r$, that is $\alpha(r)$.
\begin{figure}
\centering
\includegraphics[width=0.95\linewidth]{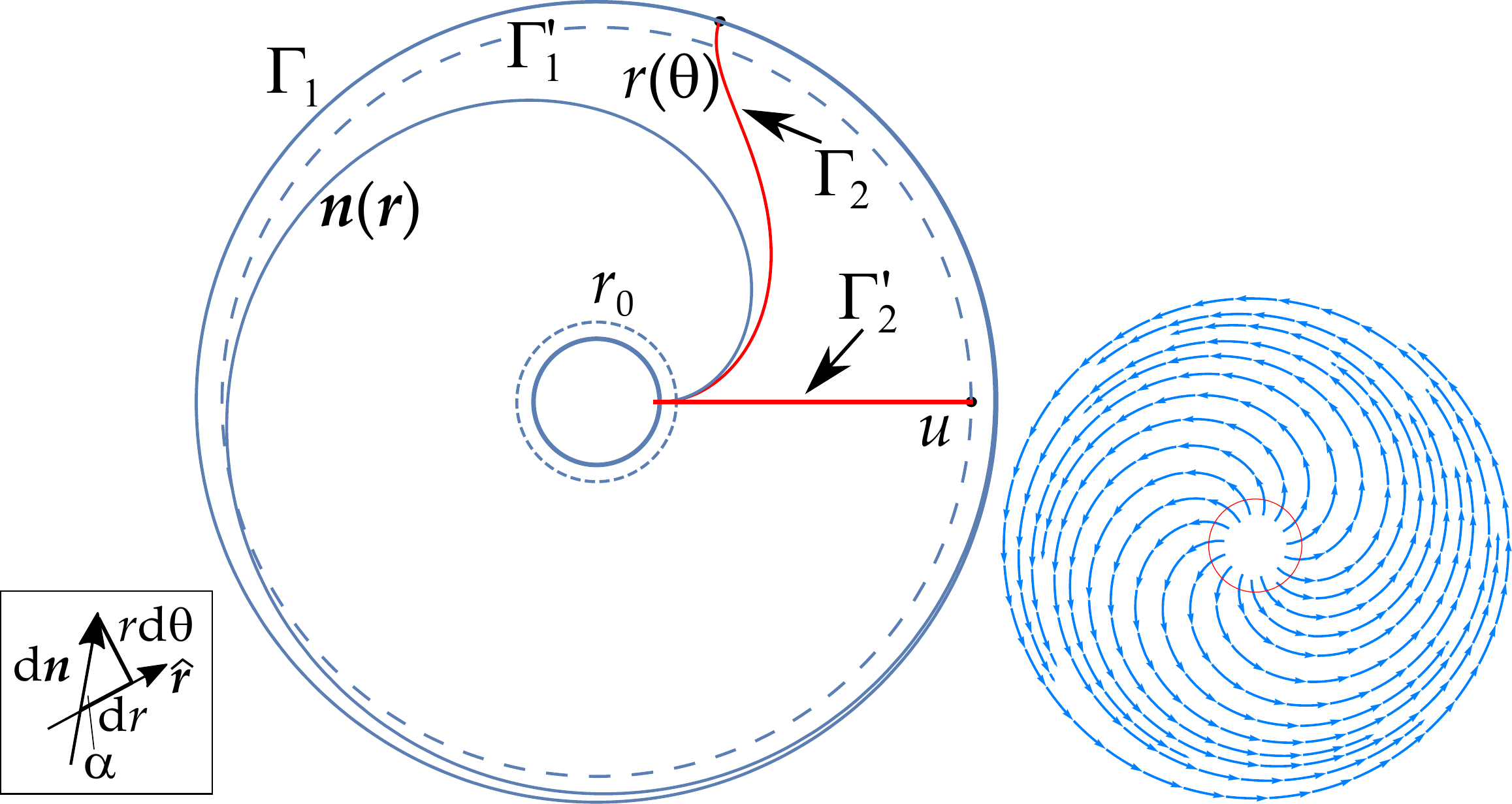}
  \caption{ Centre: An integral curve of $\n(\r)$ in the initial, flat space, for $\alpha(r) = \alpha_0 + \alpha_1\ln(r/r_0)$. The inner and outer ($\Gamma_1$) solid circles are where the director angle to the radius is $\alpha = 0$ and $\alpha  \rightarrow \pi/2$ respectively. The pattern repeats in circular bands of exponentially varying width.  A section ($\Gamma_2$) of the proto-radius' initial state integral curve $r(\theta)$ starts on the inner, full circle where $\alpha =0$ and ends on the outer. The dotted circle, $r = r_0$, is where $\alpha = \alpha_0$. The image $\Gamma_2'$ of $\Gamma_2$ is the distorted (3-D) state's new radius (the in-material distances represented here on the same diagram). It ends on the deflated circle  $\Gamma_1'$, the image of  $\Gamma_1$ at new intrinsic length $u$ from the centre, and starts a little inside the inner circle. [For $\alpha_0 = 0.2$, $\alpha_1 = 0.85$, $\lambda = 0.7$, $\nu = 2$.] Right: Director integral curves corresponding to $\alpha(r)$. Solid circle $r=r_0$. Inset square: Geometry to construct director integral curve. }
  \label{fig:proto_radii}
\end{figure}In circular coordinates and with this symmetry\cite{Mostajeran2016}, the metric tensor's components are:
\begin{align} \label{polarMetric}
a_{rr}&=\lambda^2+\left(\lambda^{-2\nu}-\lambda^2\right)\sin^2(\alpha), \nonumber \\
a_{r\theta}&=a_{\theta r}=-\frac{r}{2}\left(\lambda^{-2\nu}-\lambda^2\right)\sin(2\alpha), \nonumber \\
a_{\theta\theta}&=r^2\left[\lambda^{-2\nu}-\left(\lambda^{-2\nu}-\lambda^2\right)\sin^2(\alpha)\right].
\end{align}
The Gaussian curvature is
\begin{equation} \label{eq:Gauss}
K(r)
=\frac{\lambda^{-2}-\lambda^{2\nu}}{2}\left[\left(\alpha''+\frac{3}{r}\alpha'\right)\sin
(2\alpha)+2\alpha'^2\cos (2\alpha)\right],
\end{equation}
where $\alpha'$ is $\d \alpha(r)/\d r$.

Consider as an illustration an $\n(\r)$ allowing largely analytic illustrations in what follows: $\alpha(r) = \alpha_0 + \alpha_1\ln(r/r_0)$.
From a differential triangle, the inset box of Figure~\ref{fig:proto_radii}, of a director element $\d \n$, and its elements in the radial and $\theta$ directions, one sees the director integral curve $r(\theta)$ must obey $r\d\theta/\d r = \tan\alpha(r)$. In this illustration, $\d\alpha/\d r = \alpha_1/r$ and hence:
\begin{align}
\frac{\d\theta}{\d r} &= \frac{1}{r} \tan\alpha \equiv - \frac{1}{\alpha_1} \frac{\d}{\d r} \ln(\cos\alpha(r))\;\;
\rightarrow \theta(r) =- \frac{1}{\alpha_1} \ln(\cos\alpha(r)) \nonumber\\
\rightarrow r(\theta) &=r_0\exp\left[\frac{1}{\alpha_1}\cos^{-1}\left(\e^{-\alpha_1\theta}\right) -\alpha_0/\alpha_1\right] \label{director-integral}.
\end{align}
The origin of $\theta$ is where $\alpha = 0$,  at a radius $r_{\rm in} = r_0\exp(-\alpha_0/\alpha_1)$. The maximal $\alpha$ in this band of the director field is $\pi/2$ (and thus an azimuthal $\n$) and occurs at $r = r_0\exp((\pi/2-\alpha_0)/\alpha_1) \equiv r_{\rm in} \exp(\pi/2\alpha_1)$; see Figure~\ref{fig:proto_radii}. The pattern then repeats in radial bands. For further examples of similar periodic spiral director fields, along with their stream functions, see \cite{Mostajeran2015,Mostajeran2016}.

We now construct a parametrisation of the final, curved surface by curves that are related back to their original trajectories in the initial flat space.

\subsection{Construction of in-material circles and radii in the deformed state}
Consider two curves $\Gamma_1$ and $\Gamma_2$ in the plane that, after $\F$ acts, will remain circles and become radii respectively. We will need these elements to describe the final topography, and the transformation method between the $\Gamma$ curves and the final descriptors will be repeatedly used; see~Fig.~\ref{fig:proto_radii}.

Take a parametrisation (by $t$) of  the $\Gamma$ curves as $(r(t), \theta(t))$ that is, for convenience, unit speed and thus with a unit tangent vector $\t = (\dot{r}, \dot{\theta})$ (with $\dot{} \equiv \d/\d t$), that is $\dot{r}^2 + (r\dot{\theta})^2 = 1$. It transforms to $F\cdot (\dot{r}, \dot{\theta})$. By symmetry, the circle $\Gamma_1$ with $\dot{r}_1 =0$  must transform into an in general de/inflated circle, $\Gamma_1'$.  Demanding that the transformed tangent vectors are orthogonal (so $\Gamma_2$ generates a radius curve $\Gamma_2'$ in the sense that it meets the circle $\Gamma_1'$ perpendicularly) yields
\begin{align}
((\dot{r}_2, \dot{\theta}_2)\cdot \F^T)\cdot ( \F\cdot (\dot{r}_1, \dot{\theta}_1) )&\equiv (\dot{r}_2, \dot{\theta}_2)\cdot \a \cdot (\dot{r}_1, \dot{\theta}_1) = 0, \nonumber \\
\rightarrow \dot{r}_2 a_{r\theta} + \dot{\theta}_2 a_{\theta\theta} &= 0 \;\; \rightarrow \d r_2/\d \theta_2 = - a_{\theta\theta}/a_{r\theta} \label{transtangents}
\end{align}
on using $\F^T\cdot \F = \a$. The final result specifies the proto-radius $\Gamma_2$, that is $r_2(\theta_2)$.
With our illustrative choice of $\alpha(r)$ entering the elements of the metric, equations~(\ref{polarMetric}), we obtain explicitly:
\begin{align}
\frac{\d\theta}{\d r} &= \frac{1}{2r} \frac{\left(\lambda^{-2\nu}-\lambda^2\right)\sin 2\alpha}{\lambda^{-2\nu}-\left(\lambda^{-2\nu}-\lambda^2\right)\sin^2\alpha}  \nonumber\\
 &= - \frac{1}{2\alpha_1} \frac{\d}{\d r}\ln\left(\lambda^{-2\nu}-\left(\lambda^{-2\nu}-\lambda^2\right)\sin^2\alpha\right)  \nonumber\\
 \rightarrow \e^{-2\alpha_1\theta} &=1-\left(1-\lambda^{2(1+\nu)}\right)\sin^2\alpha \label{proto-integral-theta},
\end{align}
with $\theta = 0$  where $\alpha = 0$. Inverting eqn~(\ref{proto-integral-theta}) for $\alpha(\theta)$ and using $\alpha(r) = \alpha_0 + \alpha_1\ln(r/r_0)$ in the form of $r(\alpha)$, the proto-radius' integral curve is thus:
\be
r(\theta) = r_0 \exp\left[ \frac{1}{\alpha_1} \sin^{-1}\sqrt{\frac{1-\e^{-2\alpha_1\theta}}{1-\lambda^{2(1+\nu)}}} - \frac{\alpha_0}{\alpha_1}\right]
\ee
and is depicted in Figure~\ref{fig:proto_radii}. This would-be radius starts and ends perpendicular to circles where it is respectively along or perpendicular to the director. It remains so during deformation since at these points there is no rotational effect: The proto-radius's tangent vector at these points is along principal directions of the deformation. Between these points there is an inflectional point of maximal angle to the radial direction. When $\alpha = $const, one has a log spiral for $\n$ and the proto-radii can trivially be shown to be also (different) log spirals \cite{Mostajeran2017dragging}.

The lengths of the intrinsic circles and radii in the transformed state are not necessarily in the ratio $2\pi$ and hence there is enclosed GC that we will now calculate by ascertaining the lengths of the transformed curves $\Gamma_1'$ and $\Gamma_2'$. One can further see from Fig.~\ref{fig:proto_radii} that transformation gives material rotation, since $\Gamma_2 \rightarrow \Gamma_2'$ by an amount that depends on radial position $r$. The transformation of $\Gamma_1$ by $F$ gives a new circle $\Gamma_1'$ with a new element of length:
\begin{align}
(\d t')^2 &\equiv
\left[ (0, \dot{\theta}_1)\cdot \a \cdot (0, \dot{\theta}_1)\right](\d t)^2\; \rightarrow \d t' = \sqrt{a_{\theta\theta}/r^2}\, r \d \theta_1 \nonumber \\ l_1 &= 2 \pi r \sqrt{a_{\theta\theta}/r^2} \label{new-circumference}
\end{align}
where $l_1$ is the de/inflated circumference. Questions of new geodesics and the effect of rotations are explored more generally in \cite{Mostajeran2017dragging}.  In particular it is there demonstrated \emph{how} proto radii are rotated locally whereas circles naturally remain circles.

\subsection{Circularly symmetric director fields and surfaces of revolution}\label{forward method}
We employ a standard calculation of the GC, $K(r)$, in terms of the specification of points on a surface of revolution using projected radii and height functions $\gamma_1(u), \gamma_2(u)$, see figure~\ref{fig:surface-spec}, where the parameter $u$ is the \textit{intrinsic}, radial length from the pole of the surface.
\begin{figure}
\centering
\includegraphics[width=.5\linewidth]{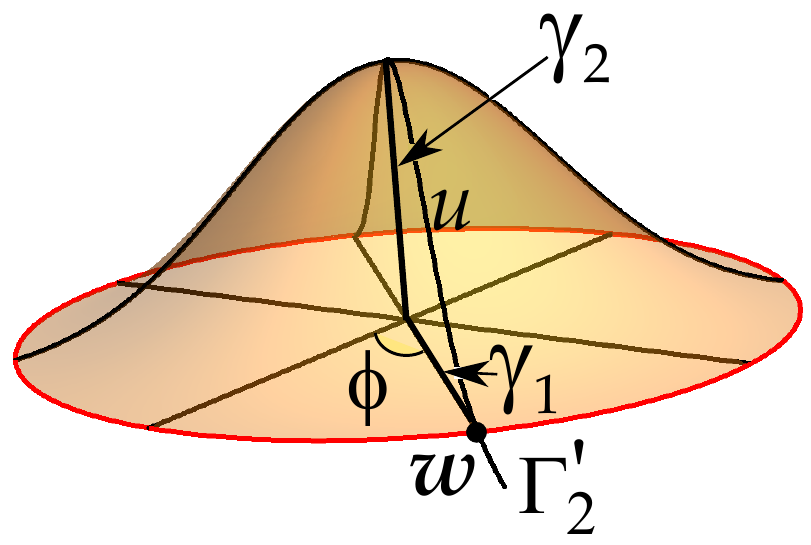}
  \caption{A point $\w$ (heavy dot) in the cylindrically-symmetric surface is specified by the radius $\gamma_1(u)$ and the height function $\gamma_2(u)$, both functions of the in-surface radial length $u$ from the apex. The usual azimuthal angle in such cylindrical coordinates is $\phi$.}
  \label{fig:surface-spec}
\end{figure}
 Comparing the $K(r)$  obtained from $\gamma_1$ and $\gamma_2$ with that obtained from the metric tensor variation, we obtain expressions for the functions $\gamma_i$.

With $c$ and $s$ shorthand for $\cos\phi$ and $\sin\phi$ respectively, the variation of the surface point $\w = (\gamma_1(u) c, \gamma_1(u) s, \gamma_2)$ in the $u$ and $\phi$ directions yields the respective unit tangents. Radially, one obtains $\t_u = \partial \w / \partial u = (\gamma_1'c,\gamma_1's,\gamma_2')/(\gamma_1'^2 + \gamma_2'^2)^{1/2}$, where $\gamma' \equiv \d\gamma/\d u$, differentiation with respect to its argument $u$. Since the unit speed representation is differentiable, then $\gamma_1'^2 + \gamma_2'^2 = 1$. The azimuthal, unit tangent is $\t_{\phi} \propto \partial \w/\partial \phi \rightarrow (-s,c,0)$. These orthogonal, unit tangents generate the unit normal to the surface $\N = \t_u \wedge \t_{\phi} = (\gamma_2' c, \gamma_2's, - \gamma_1')/(\gamma_1'^2 + \gamma_2'^2)^{1/2}$.

The rate of change of $\N$ with $\gamma_1\phi$ in the $\phi$ direction determines the associated curvature $1/R_{\phi}$:
\be
\frac{1}{R_{\phi}} = \left| \frac{1}{\gamma_1}\frac{\partial \N }{\partial \phi}\right| = \left|\frac{\gamma_2'}{\gamma_1} (-s,c,0) \right| = \frac{\gamma_2'}{\gamma_1}.
\ee
The other principal curvature, $1/R_u$, derives from the normal's in-material variation with $u$, that is $\partial \N/\partial u$. One can easily show, from the orthogonality of $\t_u$ and $\t_{\phi}$, that $1/R_u$ depends only on the variation of $\t_u$, and is $1/R_{\phi} = \left| \partial \t_u/\partial u \right|$. After a little algebra, one can show that $1/R_u = (\gamma_1''\gamma_2' -\gamma_1'\gamma_2'')/(\gamma_1'^2 + \gamma_2'^2)$. In deriving these results, we can set $\gamma_1'^2 + \gamma_2'^2 = 1$ at the end, and along the way use this constraint in its differentiated form: $\gamma_1'\gamma_1'' + \gamma_2'\gamma_2'' = 0$.

Recognising that $K = 1/(R_u R_{\phi})$, we arrive at:
\be\label{eq:surface-Gauss}
K(u) = -\gamma_1''/\gamma_1.
\ee
The form of $K$ is only deceptively simple since we have the connections between $\gamma_1'$ and $\gamma_2'$, and the constraints $0 < \left| \gamma'\right| < 1$ which considerably restrict solutions. Indeed we will find regions where we contravene this essential condition because the surfaces are not those of revolution, as we have assumed, and one has moved away from the underpinning assumptions made. Note also that $K(u)$ is a function of the in-material radius $u$ rather than of the original parameterisation radius $r$ that our other, metric-derived expression (\ref{eq:Gauss}) for $K$ depends. Therefore, except for the case of constant GC (addressed for actual examples of surfaces in \cite{aharoni2014geometry,Mostajeran2015,Mostajeran2016}), we need to connect $r$ and $u$ in order to solve equation~(\ref{eq:surface-Gauss}) with a $K(r)$ from the metric to obtain $\gamma_1$ and $\gamma_2$, and thus the surface.

The proto-radius $\Gamma_2$ tangent vector $(\dot{r}_2, \dot{\theta}_2)$ transforms under $\F$ to give a new element of length $\d u$ along $\Gamma_2'$ of $\d u = \left| \F\cdot (\dot{r}_2, \dot{\theta}_2)\right|\d t$, whence the new curve's length is:
\begin{align}
u &= \int \d t \sqrt{(\dot{r}_2, \dot{\theta}_2)\cdot \F^T \cdot \F \cdot (\dot{r}_2, \dot{\theta}_2) } 
 = \int^{r_2}_0 \d r_2' \left[ a_{rr} + 2a_{r\theta}\dot{\theta}_2/\dot{r}_2  + a_{\theta\theta} (\dot{\theta}_2/\dot{r}_2)^2 \right]^{1/2} \label{translengths}
\end{align}
on taking out $\d r_2/\d t$ and using $\F^T\cdot \F = \a$. Further, the last part of equation~(\ref{transtangents}) which defines the proto-radius $\Gamma_2$ in flat space, $\d r_2/\d \theta_2 \equiv  \dot{r}_2/\dot{\theta}_2 = - a_{\theta\theta}/a_{r\theta}$, means $u$ simplifies to:
\begin{align}
u &= \int^{r_2}_0 \d r_2' \sqrt{\det{\a}}/\sqrt{a_{\theta\theta}} =  \lambda^{1-\nu}\int^{r_2}_0 \d r_2'\, r_2'/\sqrt{a_{\theta\theta}(r_2')} \nonumber \\
 &= \lambda^{1-\nu}\int^{r_2}_0 \d r_2'/\sqrt{\lambda^{-2\nu} - (\lambda^{-2\nu} - \lambda^{2})\sin^2(\alpha(r_2'))} \label{length_result}
\end{align}
since $\det{\a} = r^2 \lambda^{2(1-\nu)}$.
A trivial example is a log spiral with $\alpha =$ const, independent of $r_2'$. One then directly obtains $ u = \lambda^{1-\nu} r_2/\sqrt{\lambda^{-2\nu} - (\lambda^{-2\nu} - \lambda^{2})\sin^2(\alpha)}$, with $u \propto r_2$, which is indeed characteristic of a cone \cite{modes2010disclinations} with $K=0$ everywhere except for a concentration of GC at its tip.

We transform variables from $u$ to $r$ (in effect $r_2$) by taking $g_1(r) \equiv \gamma_1(u(r))$ say, whence $\d \gamma_1/\d u = (\d r/\d u)( \d g_1/\d r)$ where the coefficient of $\d g_1/\d r$ can be expressed as a function of $r$:
\be \frac{\d r}{\d u}(r) = 1/(\d u/\d r) = \sqrt{\lambda^{-2\nu} - (\lambda^{-2\nu} - \lambda^{2})\sin^2(\alpha(r))}/\lambda^{1-\nu}\label{eq:drdu}.
\ee
The second derivative needed for eqn~(\ref{eq:surface-Gauss}) also follows: 
\begin{equation}
\d^2\gamma_1/\d u^2 = (\d r/\d u)^2 \d^2 g_1/\d r^2 + (\d^2 r/\d u^2) \d g_1/\d r,
\end{equation}
 with
\be \frac{\d^2 r}{\d u^2}(r) = - \alpha'(\lambda^{-2\nu} - \lambda^{2})\sin(2\alpha(r))/(2 \lambda^{2(1-\nu)}),\ee
where $\alpha' \equiv \d\alpha/\d r$. The functions $\frac{\d r}{\d u}(r)$ and $\frac{\d^2 r}{\d u^2}(r)$ enter the differential equation~(\ref{eq:surface-Gauss}) for the surface in terms of its original coordinate $r$:
\be
  \left( \frac{\d^2 r}{\d u^2}\right) g_1' + \left( \frac{\d r}{\d u}(r)\right)^2 g_1'' = - g_1 K(r).
\ee
The curvature on the right hand side is that derived from the metric variation, equation~(\ref{eq:Gauss}), and depends on $\alpha(r)$, $\alpha'(r)$ and $\alpha''(r)$, as do the coefficients of $g_1'$ and $g_1''$, the derivatives with respect to their argument, $r$, that is here $' \equiv \d/\d r$. It is this equation we now solve to illustrate a range of  shells resulting from an imprinted director distribution steering the mechanical response.

Since the original perimeter $2\pi r$ deforms to an $l_1 = 2\pi r \sqrt{a_{\theta\theta}/r^2}$, see equation~(\ref{new-circumference}), and that from Fig.~\ref{fig:surface-spec} clearly $l_1 = 2\pi \gamma_1$, then
\be
g_1(r) \equiv \gamma_1(u(r)) = r \sqrt{a_{\theta\theta}/r^2} = r \left( \lambda^{-2\nu}-\left(\lambda^{-2\nu}-\lambda^2\right)\sin^2\alpha(r) \right)^{1/2}\label{eq:g1}.
\ee
One can confirm that this $g_1(r)$ is indeed \textit{the}\footnote{One can see that, although equation~(\ref{eq:surface-Gauss}) is linear in $g_1$, one cannot take multiples of any solution for $g_1$ to obtain another since doing so upsets the geometric foundations (radii and circumferences) on which the solution was based.}
 solution of equation~(\ref{eq:surface-Gauss}). Having thus solved for $g_1$, one simply needs to solve for the height function through $\d \gamma_2/\d u = \sqrt{1 - (\d \gamma_1/\d u)^2}$ where $\d \gamma_1/\d u = (\d g_1/\d r). (\d r/\d u)$. Thus, rather in terms of $r$, we have for $g_2$:
\be \label{eq:g2-ODE}
\d g_2/\d r = \sqrt{\left( \frac{\d u}{\d r}(r)\right)^2 - \left( \frac{\d g_1}{\d r}\right)^2}.
\ee
The elements of the right hand side of equation~(\ref{eq:g2-ODE}) are all known functions of $r$, once $\alpha(r)$ is given, and the surface is specified by the solution of this first order ODE in $r$ for $g_2(r)$ and $g_1(r)$ above.

\subsection{Examples of topographies generated by spiral director patterns}\label{sect:examples}
Topography develops because of a spatially-varying deformation gradient, itself arising from a varying director field. For the forward problem, from director field to topography, the above machinery makes it straightforward to select $\alpha(r)$ fields and then find the resultant topographies, which we now illustrate by three example functions for $\alpha(r)$, with sample values of material parameters and geometry specifiers.

Surfaces must have $ (\d \gamma_1/ \d u)^2 \equiv \left(\frac{\d r}{\d u}\d g_1/ \d r\right)^2 \le 1$. Using eqn~(\ref{eq:drdu}) for $\d r/\d u$ and eqn~(\ref{eq:g1}) for $g_1(r)$, we can express this condition as:
\be
-1 \le \lambda^{-(1+\nu)} (1-(1-\lambda^{2(1+\nu)})\sin^2\alpha) - \half r \lambda^{-(1+\nu)} (1-\lambda^{2(1+\nu)}) \frac{\d}{\d r} \sin^2\alpha \le 1.
\ee
The upper bound turns out to be important, and constrains the choice of $\alpha(r)$ at $r=0$ and at a higher, and thus limiting value of $r$. At $r \rightarrow 0$ one requires:
\be
\sin^2(\alpha(0)\s{min}) = 1/(1 + \lambda^{1 + \nu})\label{eq:alpha-min}
\ee
which is a constraint recurrent in the examples below. Clearly, $\alpha(0)\s{min} > \pi/4$, with equality at $\lambda = 1$.

If the offset at the origin of $\alpha(r=0)$ is greater than $\alpha(0)\s{min} = \arcsin(1 + \lambda^{1 + \nu})^{-1/2}$, then the resulting shells have a finite point at $r=0$, that is $\d\gamma_1/\d u < 1$. Otherwise if $\alpha(r=0) =\alpha(0)\s{min}$ they are flat, with $\d\gamma_1/\d u = 1$.

\subsubsection{Linear director angle variation}\label{subsect:linear}
Consider $\alpha(r) = \alpha_0 + r$ with geometric and material parameters $\alpha_0 = 1.01$, $\lambda = 0.85$ and $\nu = 2$.
The  Gaussian curvature and the shell topography arising from this linear variation of $\alpha$ with $r$ are shown in Figure~\ref{fig:linear}.
\begin{figure}
\centering
\includegraphics[width=0.9\linewidth]{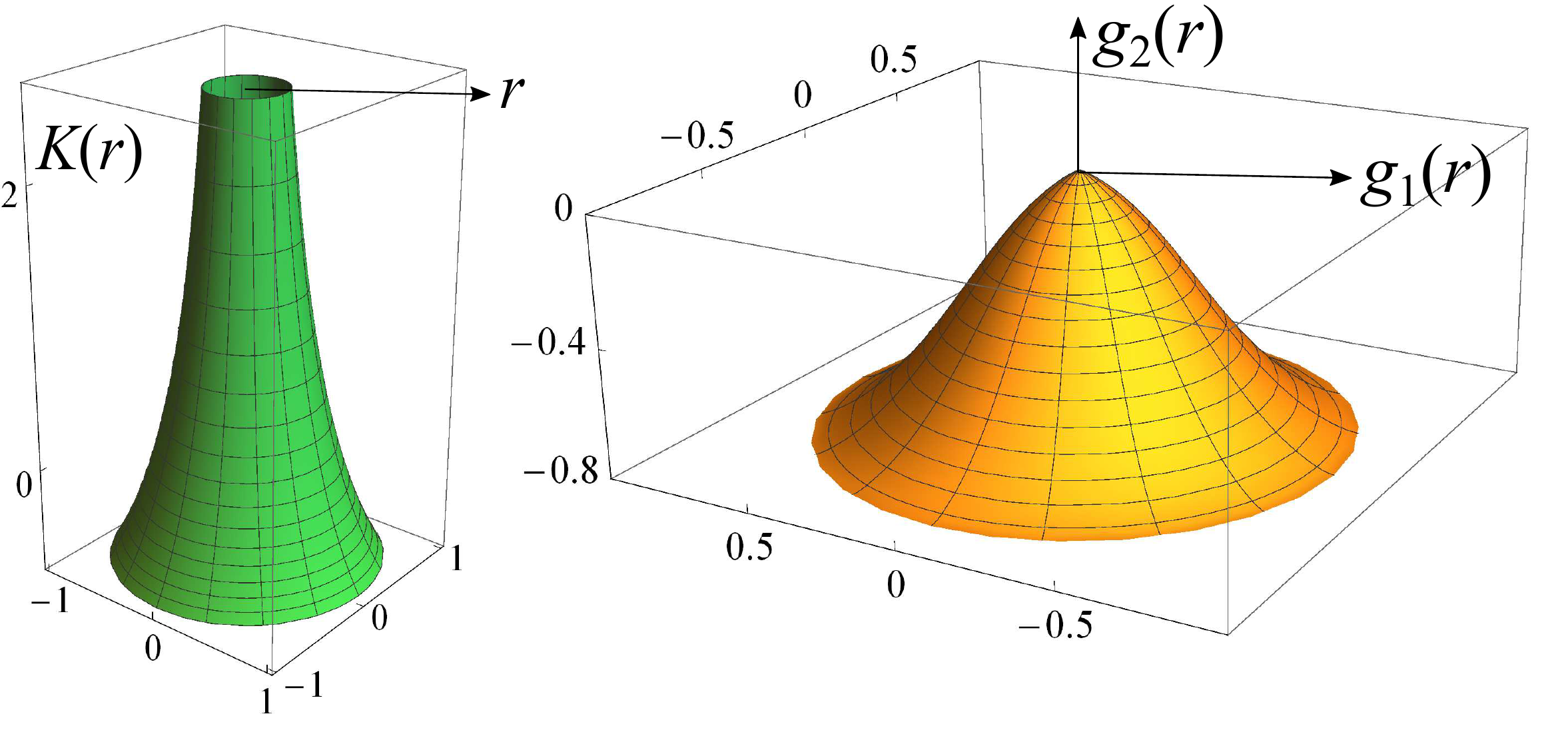}
  \caption{Calculated Gaussian curvature (left) and shell shape (right) arising from a deformation $\lambda = 0.85$, $\nu = 2$, in a spatially varying director field $\alpha(r) = \alpha_0 + r$, with $\alpha_0 = 1.01$. The GC is plotted against the radial coordinate $r$ of the initially flat, undistorted reference state.  The shell is plotted parametrically with $r$ specifying the radial coordinate $g_1(r)$ and the height function $g_2(r)$. Both plots correspond to the same region $r \in [0,1]$ of the reference state. }
  \label{fig:linear}
\end{figure}

\subsubsection{Quadratic director angle variation}\label{subsect:quadratic}
Consider  $\alpha(r) = \alpha_0 + r^2$ with geometric and material parameters $\alpha_0 = 1.25$, $\lambda = 0.51$ and $\nu= 1.5$.
The  Gaussian curvature and the shell arising from this quadratic variation of angle $\alpha$ with $r$ are shown in Figure~\ref{fig:quadratic}.
\begin{figure}
\centering
\includegraphics[width=0.8\linewidth]{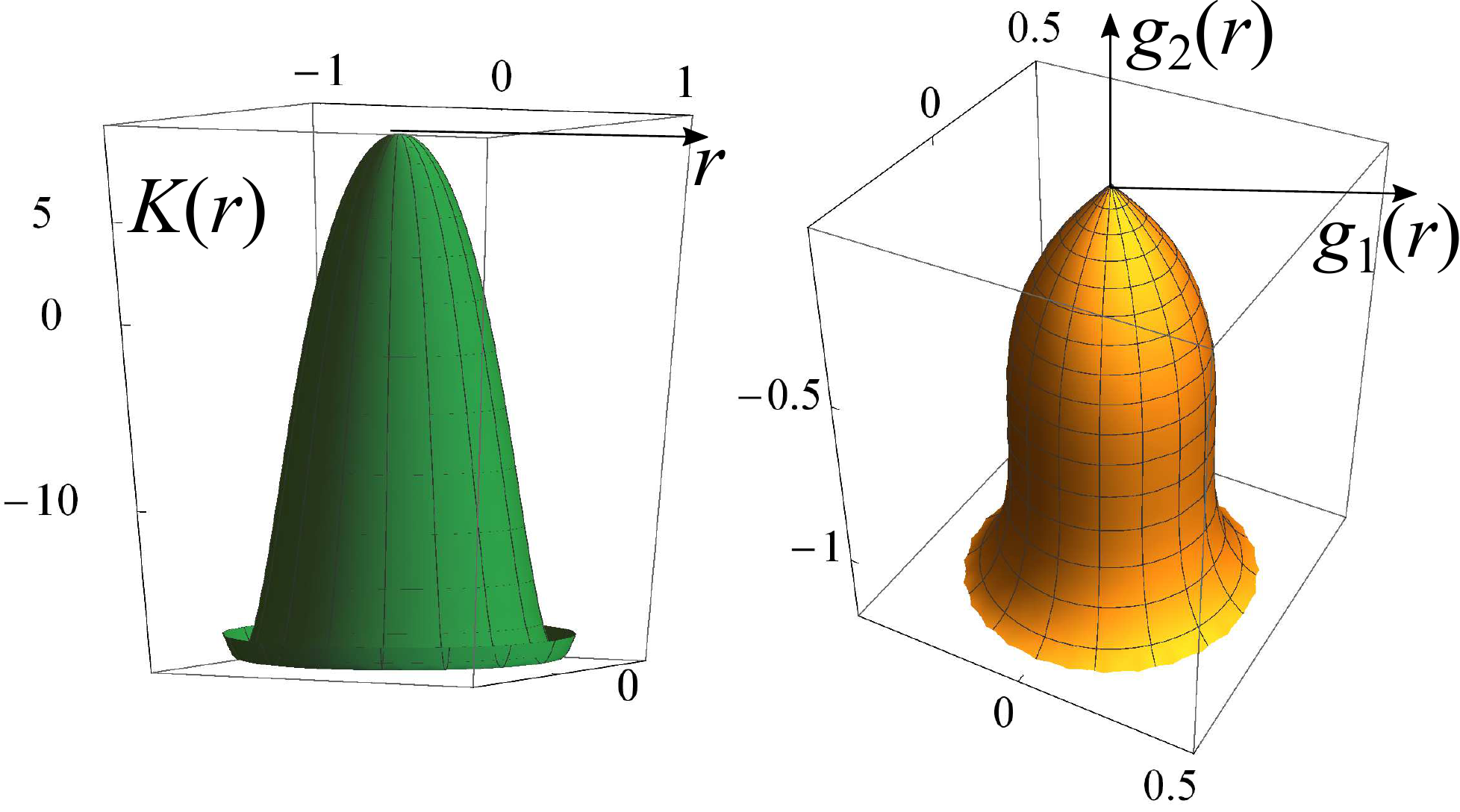}
  \caption{Calculated Gaussian curvature (left) and shell shape (right) arising from a deformation $\lambda = 0.51$, $\nu = 1.5$, in a spatially varying director field $\alpha(r) = \alpha_0 + r^2$, with $\alpha_0 = 1.25$. Coordinates and range of plots are as in Fig~\ref{fig:linear}.}
  \label{fig:quadratic}
\end{figure}
The integral curves of the director field with this $\alpha(r)$ will be shown in fig.~\ref{fig:director_field_paraboloid}.

\subsubsection{Exponential director angle variation}\label{subsect:exponential}
For an exponential director angle variation $\alpha(r) = \alpha_0 + \alpha_1\left(1- \e^{-r}\right)$, with geometric and material parameters $\alpha_0 = 0.956$, $\alpha_1 = 1.245$, $\lambda = 0.76$ and $\nu= 1.5$, one has the Gaussian curvature and shell topography shown in Figure~\ref{fig:exponential}.
\begin{figure}
\centering
\includegraphics[width=1\linewidth]{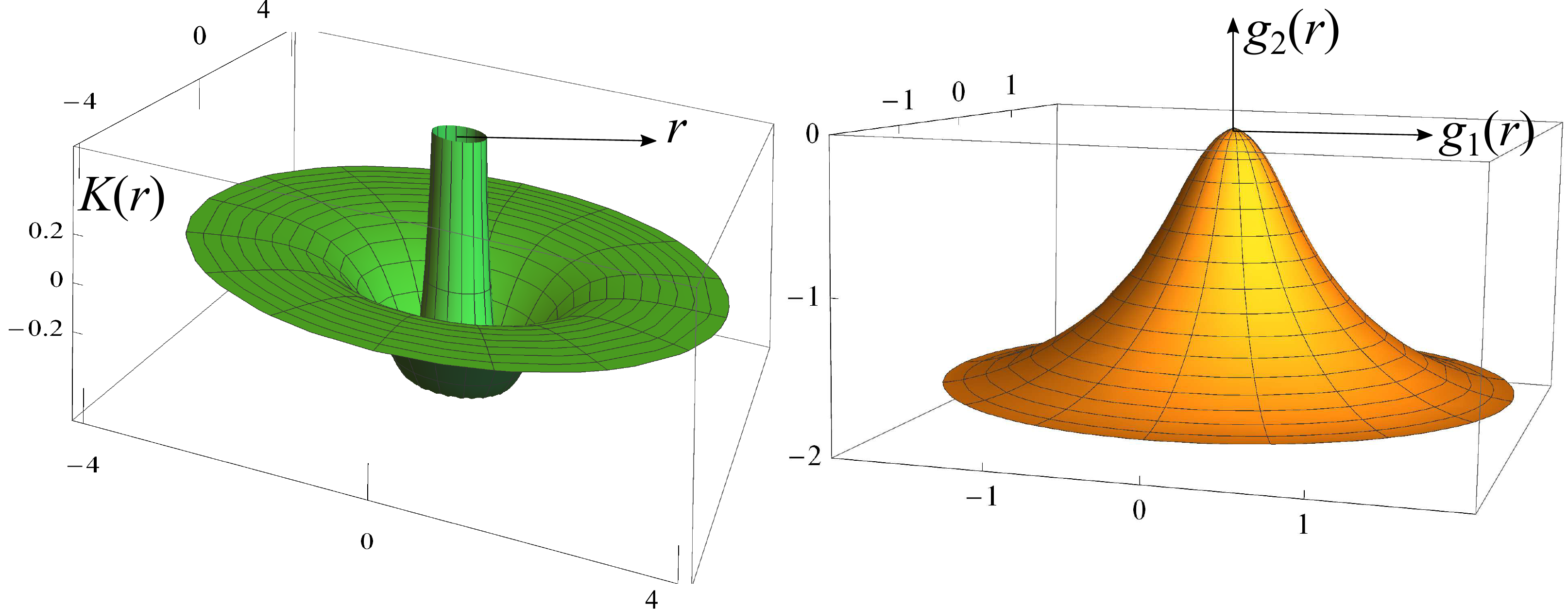}
  \caption{Calculated Gaussian curvature (left) and shell shape (right) arising from a deformation $\lambda = 0.76$, $\nu = 1.5$, in a spatially varying director field $\alpha(r) = \alpha_0 + \alpha_1\left(1- \e^{-r}\right)$, with $\alpha_0 = 0.956$ and $\alpha_1 = 1.245$. Coordinates of the plots are as in Fig~\ref{fig:linear}, but the radial range in the reference state is $r \in [0,4]$.}
  \label{fig:exponential}
\end{figure}

%

\section{The inverse problem}
The inverse challenge is to determine what director distribution $\alpha(r)$ is required to be able to generate a given circularly-symmetric shell. We have seen in the forward direction one has to at most solve a first order ODE, eqn~(\ref{eq:g2-ODE}), but the inverse problem throws up mathematical challenges that vary enormously in complexity depending on the specified shell. We give examples of where a simple, non-linear first order ODE is encountered, and solved, but in general highly non-linear integro-differential equations result.

Consider the two key defining equations for the resulting shell and its connection with the flat-space director field:
\begin{align}
\gamma_1 &= \sqrt{a_{\theta\theta}(r)} = \lambda^{-\nu} r \left(1-\beta\sin^2\alpha(r)\right)^{1/2} \label{eq:inverse-1}\\
u(r)  &= \lambda^{1-\nu} \int_0^r \d r'/ \sqrt{a_{\theta\theta}(r')/r'^2} \equiv  \lambda \int_0^r \frac{\d r'}{\left(1-\beta\sin^2\alpha(r')\right)^{1/2}} \label{eq:inverse-2}\\
 & \Rightarrow \frac{\d u(r)}{\d r} = \lambda^{1-\nu}/ \sqrt{a_{\theta\theta}(r)/r^2} \equiv \lambda^{1-\nu} r /\gamma_1 \label{eq:inverse-3}.
\end{align}
where (\ref{eq:inverse-1}) is a compressed form of eqn~(\ref{eq:g1}), and (\ref{eq:inverse-2}) is equally of eqn~(\ref{length_result}), using $\beta = 1 - \lambda^{2(1+\nu)}$. The difficulty exposed by equation~(\ref{eq:inverse-1}) is that the surface $(\gamma_1(u),\gamma_2(u))$  can be expressed parametrically in the intrinsic radial distance $u$, but the curvature is in terms of the flat, reference space coordinate $r$. Equation~(\ref{eq:inverse-2}) shows $\alpha$ is buried in an integral leading, in general, to integral equations. Equation~(\ref{eq:inverse-3}) is sometimes a route to avoiding this complexity. We first illustrate some principles arising by reconsidering the simplest example:

\subsection{Spherical caps and spindles}
The spherical cap (and spherical spindles) solved in \cite{Mostajeran2016} is a simple example of the inverse problem -- given a constant curvature $K$, what is the $\alpha(r)$ that generates it? In that case one does not have to wrestle with the connection between $u$ in the target space and $r$ in the reference state since $K$ is constant.  Taking  the expression (\ref{eq:Gauss}) with constant $K$ as an ODE for the director field yields:
\be
\alpha(r) = \half \arccos\left(-\half C(K) r^2 + c \right) \label{eq:cap_director}
\ee
(eqn~(3.17) of  \cite{Mostajeran2016}) where $C=K/(\lambda^{-2} - \lambda^{2\nu})$ and $c$ is a constant of integration. It is shown that $c \le -\frac{1- \lambda^{1 + \nu}}{1+ \lambda^{1 + \nu}}$, with equality in the case of spherical caps and inequality for spherical spindles. Considering the argument of $\arccos$ in eqn~(\ref{eq:cap_director}) must be $\ge -1$, then clearly $-1\le c$. Also, the maximal radius $r\s{m}$ is given by:
\bea
r^2\s{m} &=& 4\lambda^{-(1-\nu)}(1-\lambda^{1+\nu})/K \;\; \textrm{or} \nonumber \\
r\s{m}/R &=& 2 \sqrt{\lambda^{-(1-\nu)}(1-\lambda^{1+\nu})}\label{eq:rmax}.
\eea
We have inserted the value of $C$,  used $K = 1/R^2$ where $R$ is the radius of curvature, and we have taken $c$ appropriate to a spherical cap. There is a maximum $r$ because, in deforming a flat disc to a cap, lines of longitude  have to extend, and lines of latitude have to shrink, relative to the radii and circumferences in the flat state from which they evolve. Roughly speaking, in the extreme limit radii extend by $\lambda^{-\nu}$ and circles contract by $\lambda$ when $\alpha$ has advanced to $\pi/2$. This ratio of distortions must be sufficient to accommodate the geometric changes in trying to smooth a flat sheet around a sphere.  Eventually, the changes are just too great: for instance as one approaches the South pole, geometry requires the former to approach $\pi R$, and the latter to approach $0$, for which the $\lambda$ factors are inadequate and the cap that can be formed is limited in extent. We meet the limitation also below for catenoids and paraboloids. This limitation can be partly obviated by adapting ``petals" \cite{Priimagi_Iris_2017}, or conversely when flattening a spherical shell by partially cutting into sectors as can be left behind in an elaborately peeled orange.

What fraction of a spherical shell can be achieved for a given $\lambda$ an be directly determined: It is also shown in \cite{Mostajeran2016} that the in-surface radius of the cap, $u$ as defined above generally in eqn~(\ref{length_result}), is related to the radial distance $r$ in the reference state as:
\bea
u(r) &=& \frac{\lambda^{1-\nu}}{\sqrt{\mu_2}}\arctan\left( \frac{\sqrt{\mu_2} r}{\sqrt{\mu_1 -  \mu_2 r^2}} \right) \\
\rightarrow \tan^2\left( \phi\s{m}/2\right)&=& \frac{1-\lambda^{1+\nu}}{\lambda^{1+\nu}}\label{eq:phi-max}.
\eea
Here  $\mu_1 = \half \left[ \lambda^{-2\nu} + \lambda^2 +  c(\lambda^{-2\nu} - \lambda^2 )\right]$, $\mu_2 = \quarter \lambda^{2(1-\nu)} K$. To arrive at equation~(\ref{eq:phi-max}), we have taken $K= 1/R^2$, as well as using $C$ and $c$ for a cap, and the above expression~(\ref{eq:rmax}) for the corresponding $r\s{m}$. Here $\phi = u/R$ and is the semi opening angle for the cap. It gives a universal expression for the shape limitation explained above. For instance, $\lambda \rightarrow 0$ would be required to make a sphere, $\phi\s{m} = \pi$. For a hemisphere, $\phi\s{m} = \pi/2$, one would require $\lambda\s{hs} = (2/3)^{1/(1+\nu)} \rightarrow \lambda\s{hs} = .85$ for the value $\nu = 1.5$ considered in the examples above.

\subsection{The director field required for spatially-varying  shell curvature}\label{ssect:inverse}
To best illustrate the method of director field determination given a shell shape, we explore cases that admit of a complete, analytical solution. We thereby explore the limits of an $\alpha$ field specifying surfaces of revolution and the geometric constraints of radii and perimeters changing in very different ways.

Surfaces of  revolution, where the in-material radius $u$ and the radius $\gamma_1$  are related as below, can be simply analysed:
\bea u &=& f(\gamma_1) \label{eq:inverse-4} \; \rightarrow  \\
\lambda \int_0^r \frac{\d r'}{\left(1 - \beta \sin^2 \alpha(r') \right)^{1/2} } &=& f \left( r \lambda^{-\nu} \sqrt{1-\beta \sin^2\alpha(r)} \right) \label{eq:inverse-general}.
\eea Again $\beta = 1 - \lambda^{2(1+\nu)}$.
In general, forms other than eqn~(\ref{eq:inverse-4}) lead to non-linear integral equations for $\sin^2\alpha$. In $f$, the argument $\gamma_1$ from eqn~(\ref{eq:inverse-1}) has been re-written as $ r \sqrt{a_{\theta\theta}(r)/r^2}$ to make the argument of the surd only depend on $r$ through the dependent variable $\alpha$.
The integral equation nature of eqn~(\ref{eq:inverse-general}) can be eliminated by differentiation with respect to $r$ which, after a little re-arrangement, yields a trivial, first order ODE which we can re-write as:
\be
\frac{\d v}{\d w} v f(v)= \lambda^{1+\nu} w .
\ee
Here $w = \lambda^{-\nu} r$ is the scaled, reference state radial length and 
\begin{equation}
v \equiv \gamma_1 = w \sqrt{1 - \beta \sin^2\alpha(w)}.
\end{equation}
Quadrature gives the relation
\be
h(v) \equiv v f(v) -\int_0^v f(v')\d v' = \half \lambda^{1+\nu} w^2 \label{eq:f-sol}.
\ee
We take $v=0$ at $w=0$ quite naturally since they involve intrinsic and extrinsic radii, but below in an example generalise to where this is not satisfied, that is with a more general constant of integration allowable when shells have a central disc excised.

It is generally easier to consider $w^2$ as a function of $v$:
\be
w^2  = \frac{2}{\lambda^{1+\nu}} h(v) \equiv \frac{2}{\lambda^{1+\nu}}\left(v f(v) -\int_0^v f(v')\d v'\right)  \label{eq:w-sol}.
\ee
The director field required to generate a shell with a given $\gamma_1(u)$ (and $\gamma_2$ from $\gamma_2' = \sqrt{1-\gamma_1'^2}$ ) is then specified by inverting the definition of $v$ to yield $\sin^2\alpha$ and employing eqn~(\ref{eq:w-sol}) for $w^2(v)$ in that inversion:
\bea
\alpha(v) &=& \sin^{-1}\left[\frac{1}{\beta}\left(1-(v/w)^2\right)\right]^{1/2} =  \sin^{-1}\left[\frac{1}{\beta}\left(1-\frac{\lambda^{1+\nu}v^2}{2 h(v)} \right) \right]^{1/2} \label{eq:para1}\\
w(v) &=& \sqrt{\frac{2}{\lambda^{1+\nu}}h(v)}\label{eq:para2}.
\eea

\noindent \textit{Conditions on the director field.}\\
From the first part of eqn~(\ref{eq:para1}), we see that
\be
(v/w)^2 \ge 1-\beta = \lambda^{2(1+\nu)}
\ee
in order that $\sin^2\alpha \le 1$. One also requires $(v/w)^2 \le 1$ for $\sin^2\alpha \ge 0$. This condition is most seriously tested at $w \rightarrow 0$ where one can expand in the right hand side of eqn~(\ref{eq:w-sol}) if $f$ does not specify a surface with a singular tip:
\be w^2_{w\rightarrow 0} \sim (f_0'/\lambda^{1 + \nu}) v^2 \;\;  \Rightarrow \; (v/w)^2_{w\rightarrow 0}  \sim f_0'/\lambda^{1 + \nu} \label{eq:limit}
\ee
where $f_0' \equiv \d f/\d v |_{v \rightarrow 0}$ is the derivative at the shell centre. From eqn~(\ref{eq:para1}) we extract the required $\alpha$ behaviour around what will become the tip as:
\be \sin^2\alpha_0 = \frac{1}{\beta} \left( 1 - \lambda^{1 + \nu}/f_0' \right) \equiv \frac{1 - \lambda^{1 + \nu}/f_0'}{1 - \lambda^{1 + \nu}}\frac{1}{1 + \lambda^{1 + \nu}}\label{eq:centre}.
\ee
The character of the apex of the shell determines the expansion of $f$:
\be u = f(\gamma_1) = f_0 + f_0' \gamma_1 + \dots \ee
and hence the initial $\alpha$. Clearly $f_0 = 0$ (no in-plane radial length when $\gamma_1 = 0$).

Shells flat at their apex have $f_0' = 1$, that is $u \sim \gamma_1$ as $u, \gamma_1 \rightarrow 0$.  We then see from eqn~(\ref{eq:centre}) the familiar result $\sin^2\alpha_0 = 1/(1 + \lambda^{1+\nu})$, that is eqn~(\ref{eq:alpha-min}).

Peaked shells have $f_0' > 1$. Simple geometry of the cone fitted to the very tip of the shell shows that $ u =\gamma_1/\sin\phi$ and thus  $f_0' = 1/\sin\phi$, where $\phi$ is the fitting cone's opening angle. Eqn~(\ref{eq:centre}), with $f_0' > 1$, clearly now requires a larger $\alpha_0$ than that leading to centrally flat shells, for the same $\lambda$.

The maximum $\alpha_0$ is $\pi/2$, whereupon from eqn~(\ref{eq:centre}) one has $f_0' = 1/\lambda^{1+\nu}$, or otherwise expressed $\sin\phi = \lambda^{1+\nu}$, which is the familiar result for a cone deriving from circular director field where $\alpha = \pi/2$, independently of $r$ .

\subsection{The director field required for a catenary of revolution}
An example of the general method above is that of a catenary of revolution (revolved about its symmetry axis). It can be specified by $\gamma_1,\gamma_2$ such that $f(v) = \sinh(v)$, that is
\be u = \sinh(\gamma_1)\label{eq:inverse-catenary}
\ee and $\gamma_2 = \cosh(\gamma_1)$.  Now solution as above gives the relation
\be
h(v) \equiv v \sinh(v) - \cosh(v) + c = \half \lambda^{1+\nu} w^2 \label{eq:catenary-sol}
\ee
with $c$ a constant of integration, the size of which is critical to the character of the solutions $v(w)$ and hence to the angle $\alpha$ that the director makes with the radial direction in the reference state. Following the procedure leading to eqns~(\ref{eq:para1}) and (\ref{eq:para2}), we show $\alpha(w)$, parametrically in $v$, in figure~\ref{fig:catenoid}(a).

The l.h.s. of eqn.~(\ref{eq:catenary-sol}), that is $h(v)$, is monotonic and, for small $v$, is:
\be
h(v\rightarrow 0) \sim -1 + c + \half v^2 + \eighth  v^4 + \dots
\ee
See figure~\ref{fig:catenoid}(a) for the three qualitatively different solutions for $\alpha$:
\begin{figure}
\centering
\includegraphics[width=0.95\linewidth]{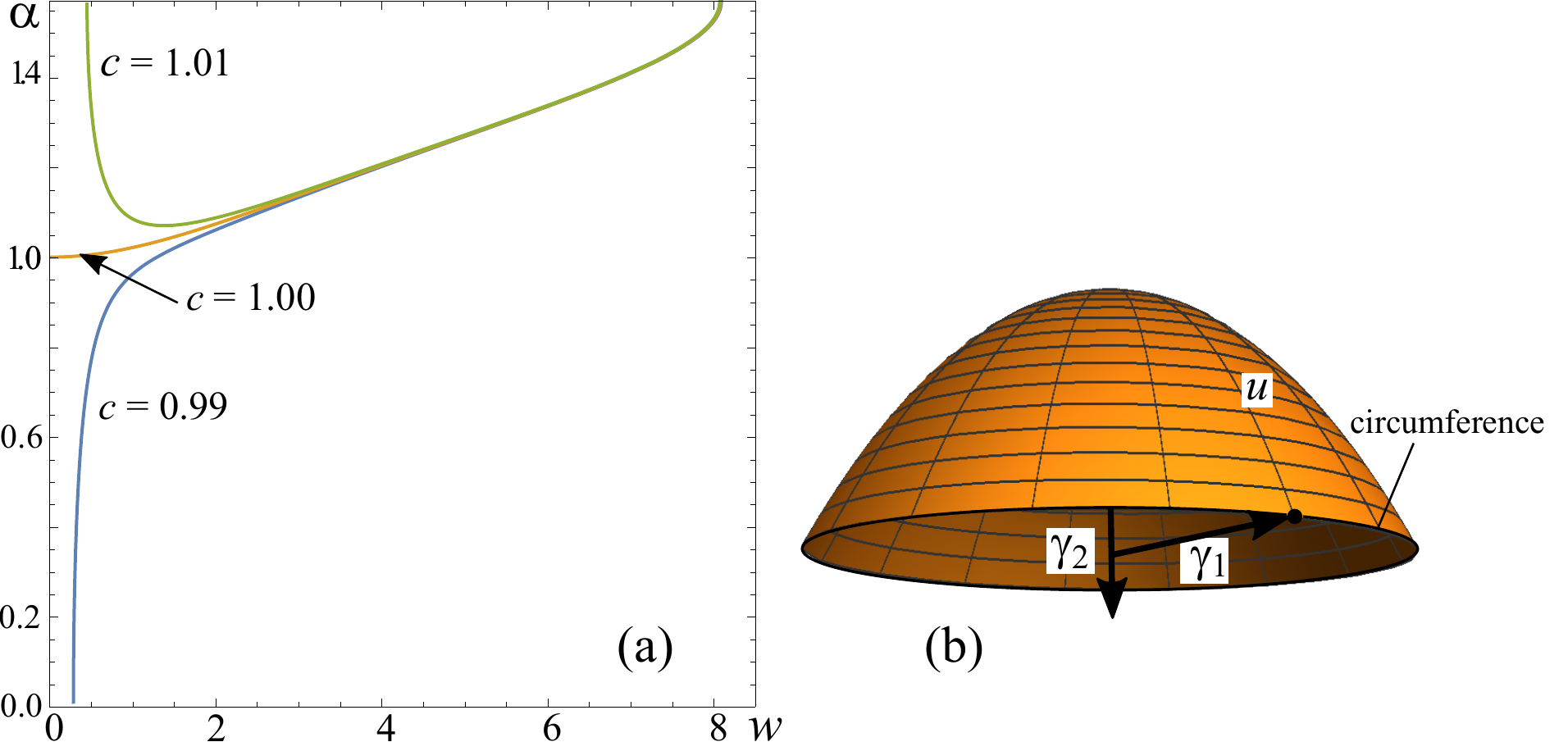}
  \caption{(a) Calculated director angle $\alpha(w)$ with respect to a radius in the reference state in order to generate a target state catenary of revolution. Radii in the reference state, $r$, are scaled by the characteristic length of the catenary, and further as $w = \lambda^{-\nu} r$. This illustration takes a deformation $\lambda = 0.7$ with $\nu = 1.5$. The curves are labelled by the value taken for the constant $c$ of integration in eqn~(\ref{eq:catenary-sol}).  (b) The in-material radial length $u$ in a (inverted) catenoid becomes exponentially longer than the embedded radius $\gamma_1$, a difference that eventually cannot be accommodated by a change in their ratio of $\lambda^{1+\nu}$ induced by heat or light on even a director field with $\alpha = \pi/2$ (which must be the director angle at the outer limit of solutions).}
  \label{fig:catenoid}
\end{figure}
For $c < 1$, there is a region of small $v$ for which $f(v) < 0$ and there are clearly no solutions for eqn.~(\ref{eq:catenary-sol}). For $c > 1$, there is a tip region of small $w$ where the right hand side, $\half \lambda^{1 + \nu} w^2$, is less than the finite limiting value $f(v\sim 0) = c-1$ of the left hand side, and again there are no solutions around the tip of the catenoid. These failures are indicative of the $\alpha(w)$ stemming from the choice of $c$ failing to generate a surface of revolution. See below where we show the choice $c=1$ avoids this failure.

For any $c$, at large $v$ we have $f(v) \sim \half v \e^v$ and eventually this exponential increase in $f(v)$ again renders solutions unattainable, but for the different geometrical reason we saw in the limitation on spherical caps:
Figure~\ref{fig:catenoid}(b) shows the in-material radius of length $u$ corresponding to the embedded radius $\gamma_1$. In the reference state this circumference was originally $2\pi r$ and is now $2\pi \gamma_1$. For large enough $r$, for a catenoid the length $u$ is exponentially larger than $\gamma_1$ and hence exponentially larger than the original circumference to which it relates. The ratio of the two has changed by $\lambda/\lambda^{-\nu} = \lambda^{1+\nu}$ and, for large $r$, this transformation ratio is not large enough to accommodate the geometry of a catenoid -- solutions are no longer possible; see figure~\ref{fig:catenoid}(a). However, the more $\lambda$ deviates  below $\lambda = 1$, the larger the range of radii leading to surfaces of revolution, as seen above for the spherical cap.

The case of $c=1$ is most interesting  since then solutions obtain down to $w=0$ (but still fail at large $w$ for the reasons given above): Around $v \sim 0$, eqn~(\ref{eq:catenary-sol}) reduces to
\be
\half v^2 + \dots \equiv \half w^2 (1 - \beta \sin^2\alpha(w)) + \dots  = \half \lambda^{1+\nu} w^2\label{eq:c1expand},
\ee
concentrating on $v^2$ terms and using the definition $v^2 = w^2 (1 - \beta \sin^2\alpha(w))$. Then as $w \rightarrow 0$, one has the limiting (minimal) $\alpha(0)\s{min}$ given before at eqn~(\ref{eq:alpha-min}), $\alpha(0)\s{min} = \sin^{-1} (1/\sqrt{1+\lambda^{1 + \nu}})$.

 Away from $w=0$, the angle $\alpha$ grows to its maximum, $\pi/2$, which it attains on the outer boundary of the reference state domain that can support formation of catenoids by contraction along a director field.  It is such an azimuthal $\n$ that gives the maximal circumferential contraction, beyond which the disparity of intrinsic and embedded radii can no longer be supported.

\subsection{The director field required for a paraboloid}\label{ssect:paraboloid}
We turn now to an important class of surfaces, paraboloids, that are susceptible to a variation of the general method stemming from $u = f(\gamma_1)$. Further we emphasise an important general point about the inverse problem, that the director distribution, once chosen for a particular $\lambda = \Lambda$, say, produces different surfaces for all other $\lambda \ne \Lambda$. One is able to make general observations about the $\lambda$-surfaces arising from the $\Lambda$ distribution of director, $\alpha_{\Lambda}(r)$ say. We also show that it is possible to anchor paraboloids in their reference state support.

For the paraboloid formed from rotating $\gamma_2 = \half a \gamma_1^2$ about $\gamma_1 = 0$, one has:
\be
\d u = \d \gamma_1 \sqrt{1 + (\d \gamma_2/\d\gamma_1)^2} = \d \gamma_1 \sqrt{1 + a^2\gamma_1^2}\label{eq:parabola}.
\ee
Differentiating w.r.t. $r$, using eqn~(\ref{eq:inverse-3}) for $\d u/\d r$, and replacing $w = \lambda^{-\nu} r$ and now $v = a \gamma_1$, one obtains
\bea
a^2\lambda^{1+\nu}w &=& v \sqrt{1 + v^2}\d v/\d w \nonumber \\
\rightarrow  \half a^2\lambda^{1+\nu}w^2 &=& \third (1 + v^2)^{3/2} - \third \label{eq:parabola-eq}
\eea
where the constant of integration, $-\third$, ensures $w=0, v = 0$. Rearrangement, and restoring $v^2 = a^2w^2(1-\beta \sin^2\alpha)$ yields an equation for $\sin\alpha$:
\be
\sin^2\alpha_{\lambda}(w) = \frac{1}{\beta_{\lambda}} \left( 1 - \frac{1}{a^2w^2}\left[ \left( 1 + \threehalf \lambda^{1+\nu} a^2 w^2 \right)^{2/3} - 1\right]\right)\label{eq:paraboloid-alpha}
\ee
where the subscript $_{\lambda}$ on the $\alpha(w)$ and the $\beta$ reminds us that this $\alpha(w)$ only yields a paraboloid for this particular $\lambda$. Expansion of eqn~(\ref{eq:paraboloid-alpha}) for $w \rightarrow 0$ yields $\sin^2\alpha_{\lambda}(w \sim 0) = 1/(1 + \lambda^{1+\nu}) + O(w^2)$, holding,  as we have argued above, for all shells flat at their apex. We show the pattern in fig.~\ref{fig:director_field_paraboloid} for $\lambda = 0.8$, $\nu = 2$, for the range $\alpha(0)$ ($=54^{\circ}$ for these $\lambda$ and $\nu$ values) to the limit of surfaces of revolution, $\alpha = \pi/2$.
\begin{figure}
\centering
\includegraphics[width=0.85\linewidth]{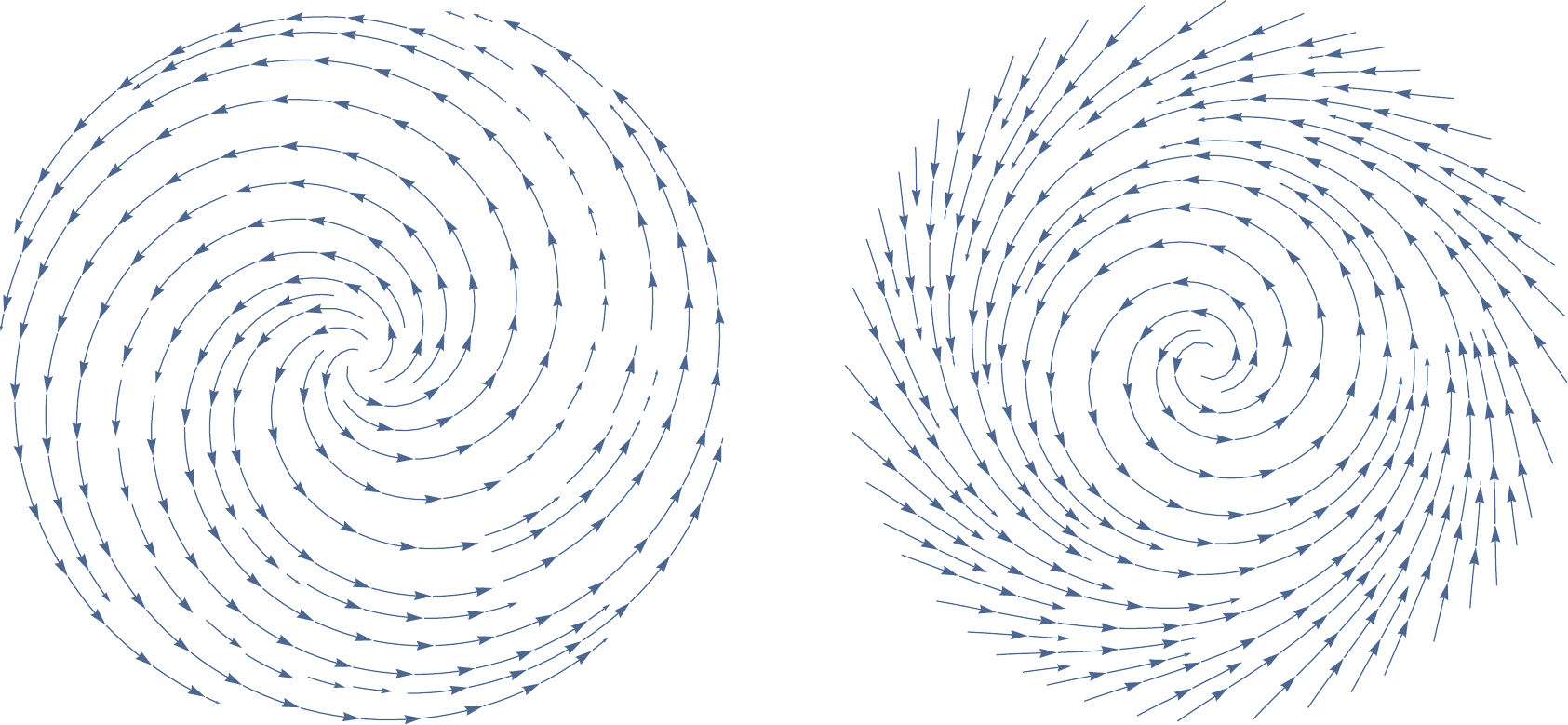}
  \caption{Left: The director field $\alpha(ar)$ required to yield a paraboloid of revolution $\gamma_2 = \half a \gamma_1^2$ when the deformation is $\lambda = 0.8$, with $\nu = 2$. Right: For comparison the integral curves for $\alpha(r) = \alpha_0 + r^2$ with $\alpha_0 = 1.25$ as employed in the forward problem shown in section~\ref{sect:examples}(\ref{subsect:quadratic}); see fig.~\ref{fig:quadratic}.}
  \label{fig:director_field_paraboloid}
\end{figure}

For increasing $w$, the director angle increases until $\sin^2\alpha_{\lambda} = 1$, at which $w$ we cease to be able to generate a shell of revolution. With this value of $\sin^2\alpha_{\lambda} $, eqn~(\ref{eq:paraboloid-alpha}) yields the condition for the maximal radius:
\be
 1+\lambda^2a^2 r^2=  \left( 1 + \threehalf \lambda^{1-\nu} a^2 r^2 \right)^{2/3}\label{eq:paraboloid-max}
\ee
where we return to $r = w \lambda^{\nu}$. This expression clearly shows that the natural reduced length in the reference space is $ar$. The reasons for the failure to realise a shell of revolution are precisely those discussed below eqn~(\ref{eq:rmax}) for spherical caps, and above eqn~(\ref{eq:c1expand}) for catenoids.

For a given $\alpha_{\Lambda}$ pattern designed to produce a paraboloid when $\lambda=\Lambda < 1$, we inquire of the sequence of shapes generated as $\lambda = 1 \rightarrow \lambda = \Lambda$ (and possibly $\lambda < \Lambda$). We thus have to return to solving the forward problem, since $\alpha(r)$ is now given, where:
\be
g_1(r) = r \lambda^{-\nu} (1 - \beta_{\lambda}\sin^2\alpha_{\Lambda}(r))^{1/2} \label{eq:g1lambda}.
\ee
For clarity, $\beta_{\lambda} = 1 - \lambda^{2(1+\nu)}$ has  its $\lambda$ dependence flagged, and the $\alpha$ field relates to $\Lambda$. To determine the shell parametrically in $r$, we need in addition to the embedded radius $g_1(r)$ the height function $g_2(r)$ given from the differential equation~(\ref{eq:g2-ODE}) where one needs care with $\d u/\d r = \lambda/\sqrt{1 - \beta_{\lambda}\sin^2\alpha_{\Lambda}(r)}$ which depends on $\lambda$ and has a memory of $\Lambda$ through $\alpha_{\Lambda}$ from eqn~(\ref{eq:paraboloid-alpha}) with $\lambda = \Lambda$. The differential equation for $g_2$ does not appear to admit of analytic solution, but is straightforward to solve numerically.

We can however analytically determine that for $\Lambda < \lambda < 1$ a sequence of peaked shells is explored, terminating in a paraboloid: The derivative $\frac{\d \gamma_1}{\d u}(r)$, considered as a function of the reference space $r$, must be bounded as $-1 \le \frac{\d \gamma_1}{\d u} \le 1$. Take $\frac{\d \gamma_1}{\d u}(r) = (\d r/\d u)(\d g_1/\d r) =  \frac{1}{\lambda}\sqrt{1 - \beta_{\lambda}\sin^2\alpha_{\Lambda}(r)} (\d g_1/\d r)$, and take eqn~(\ref{eq:g1lambda}) to evaluate the last derivative.  One then obtains for the condition that shells of revolution exist: $-1 \le \lambda^{-(1+\nu)} (1 - \beta_{\lambda}\sin^2\alpha_{\Lambda}(r)) \le 1$ around $r = 0$. We have used the next term in the expansion of eqn~(\ref{eq:paraboloid-alpha}), which is $O(w^2)$, to determine that  $\frac{\d}{\d r}\sin^2\alpha_{\Lambda}(r)|_{r\rightarrow 0} = 0$, and we further employ $\sin^2\alpha_{\Lambda}( 0) = 1/(1 + \Lambda^{1+\nu})$. After some rearrangement we obtain the condition $0\le \left(\Lambda/\lambda\right)^{1 + \nu} + \lambda^{1 + \nu}\le   1 + \Lambda^{1 + \nu}$ which is clearly satisfied\footnote{since it is equivalent to $\left( \sqrt{\lambda^{1 + \nu}} - \sqrt{\Lambda^{1 + \nu}} \right)^2 \le \left( 1 - \sqrt{\Lambda^{1 + \nu}} \right)^2$.} for $\Lambda < \lambda < 1$. This sequence of shells exists and all have, until $\lambda = \Lambda$, that $\frac{\d \gamma_1}{\d u} < 1$. They accordingly have a finite slope at $r=0$, that is they are pointed until $\lambda = \Lambda$.

\subsubsection{Fixed boundaries for mounting the paraboloid}
Invariant radii are important for anchoring for applications; see the discussion in \cite{Mostajeran2016}. From eqn~(\ref{eq:inverse-1}), the condition $\gamma_1(r) = r$ for the reference space radius $r$ to be identical to the target space embedded radius $\gamma_1$ is:
\be
\lambda^{-\nu}\left(1 - \beta_{\lambda}\sin^2\alpha_{\Lambda}(w) \right) = 1.
\ee
This condition determines the scaled radius $w$ where the non-paraboloid shells at $\lambda > \Lambda$ have natural anchoring, that is, their outer perimeter is unchanged from its value in the reference state and so it can be joined without mismatch to an inert outer region. The condition is quite simple when $\lambda = \Lambda$ where we have the actual parabola:
\be
1 + (aw)^2 \Lambda^{2\nu} = \left( 1 +  \threehalf (aw)^2 \Lambda^{1+\nu}\right)^{2/3}.
\ee
Apart from the trivial solution at $w=0$, there is another solution at finite $aw$ if $\nu > 1$.  [One sees this from expanding around $(aw)^2 \sim 0$ to see that the r.h.s dominates, whereas the l.h.s. is dominant at large $(aw)^2$.] One can check whether anchoring occurs within attainable values of $\sin^2\alpha$ by rewriting $\gamma_1(r) = r$ as $aw\lambda^\nu = v$, squaring and substituting for $v^2 = (aw)^2(1 - \beta\sin^2\alpha)$ whereupon $\sin^2\alpha = (1-\lambda^{2\nu})/(1 - \lambda^{2(1+\nu)}) < 1$ as required at the point of anchoring.

 A reasonable estimate of the anchoring radius can be achieved by assuming it occurs such that $ \threehalf \lambda^{1+\nu} a^2 w^2 \gg 1$ so that on both sides. of eqn~(\ref{eq:paraboloid-max}) the 1 terms can be neglected, whereupon $(aw)^2 \approx (\threehalf)^2 \Lambda^{2(1-2\nu)}$. \textit{Post hoc} justification of the neglect of the 1 terms, at for instance $\Lambda = 0.8$ and $\nu = \half$, can be directly tested and holds well. Thus anchoring is possible with an outer radius of $w = \frac{1}{a}\threehalf\Lambda^{1-2\nu}$.

\section{Discussion}
We have given explicit routes via which one can find  (i) the shape of revolution resulting from the action of light or heat on a given circularly symmetric director field in an initially flat plate, and (ii) the director field needed in an initially flat reference state in order to generate a specified shape of revolution. In both cases, the domain of the reference state that will generate a shape of revolution is limited by an important geometric factor that comes into conflict with the limits on how much a solid can differentially deform parallel and perpendicular to its director. These insights should inform future programmes of the limitations for developing topography for non-isometric applications such as strong actuation, micro-mechanical components and machines (gates, valves, pumps and lifters). It also suggests future strategies for obviating these limitations, for instance excision of regions of the reference state, and using piecewise continuous director distributions rather than simple functions as employed here. Practical examples for anchoring perimeters are given as illustrations of a general strategy that will be needed for practical applications. We return to difficulties and advantages within this framework.

\bibliographystyle{rspa}
\bibliography{references}

\providecommand{\noopsort}[1]{}\providecommand{\singleletter}[1]{#1}%
\begin{thebibliography}{10}
\expandafter\ifx\csname urlstyle\endcsname\relax
  \providecommand{\doi}[1]{doi:\discretionary{}{}{}#1}\else
  \providecommand{\doi}{doi:\discretionary{}{}{}\begingroup
  \urlstyle{rm}\Url}\fi

\bibitem{modes2010disclinations}
Modes, C.~D., Bhattacharya, K. \& Warner, M., 2010 Disclination-mediated
  thermo-optical response in nematic glass sheets.
\newblock \emph{Physical Review E} \textbf{81}, 060701(R).
\newblock (\doi{10.1103/PhysRevE.81.060701}).

\bibitem{modes2011gaussian}
Modes, C., Bhattacharya, K. \& Warner, M., 2011 Gaussian curvature from flat
  elastica sheets.
\newblock \emph{Proceedings of the Royal Society A: Mathematical, Physical and
  Engineering Science} \textbf{467}, 1121--1140.
\newblock (\doi{10.1098/rspa.2010.0352}).

\bibitem{aharoni2014geometry}
Aharoni, H., Sharon, E. \& Kupferman, R., 2014 Geometry of thin nematic
  elastomer sheets.
\newblock \emph{Physical review letters} \textbf{113}, 257801.
\newblock (\doi{10.1103/PhysRevLett.113.257801}).

\bibitem{Mostajeran2015}
Mostajeran, C., 2015 Curvature generation in nematic surfaces.
\newblock \emph{Phys. Rev. E} \textbf{91}, 062405.
\newblock (\doi{10.1103/PhysRevE.91.062405}).

\bibitem{Mostajeran2016}
Mostajeran, C., Warner, M., Ware, T.~H. \& White, T.~J., 2016 Encoding gaussian
  curvature in glassy and elastomeric liquid crystal solids.
\newblock \emph{Proceedings of the Royal Society A: Mathematical, Physical and
  Engineering Science} \textbf{472}, 20160112.
\newblock (\doi{10.1098/rspa.2016.0112}).

\bibitem{klein2007shaping}
Klein, Y., Efrati, E. \& Sharon, E., 2007 Shaping of elastic sheets by
  prescription of non-euclidean metrics.
\newblock \emph{Science} \textbf{315}, 1116--1120.
\newblock (\doi{10.1126/science.1135994}).

\bibitem{kim2012designing}
Kim, J., Hanna, J.~A., Byun, M., Santangelo, C.~D. \& Hayward, R.~C., 2012
  Designing responsive buckled surfaces by halftone gel lithography.
\newblock \emph{Science} \textbf{335}, 1201--1205.
\newblock (\doi{10.1126/science.1215309}).

\bibitem{dervaux2008morphogenesis}
Dervaux, J. \& Ben~Amar, M., 2008 Morphogenesis of growing soft tissues.
\newblock \emph{Physical Review Letters} \textbf{101}, 068101.
\newblock (\doi{10.1103/PhysRevLett.101.068101}).

\bibitem{de2012engineering}
de~Haan, L.~T., S{\'a}nchez-Somolinos, C., Bastiaansen, C.~M., Schenning, A.~P.
  \& Broer, D.~J., 2012 Engineering of complex order and the macroscopic
  deformation of liquid crystal polymer networks.
\newblock \emph{Angewandte Chemie International Edition} \textbf{51},
  12469--12472.
\newblock (\doi{10.1002/anie.201205964}).

\bibitem{ware2015programmable}
Ware, T.~H., Perry, Z.~P., Middleton, C.~M., Iacono, S.~T. \& White, T.~J.,
  2015 Programmable liquid crystal elastomers prepared by thiol--ene
  photopolymerization.
\newblock \emph{ACS Macro Letters} \textbf{4}, 942--946.
\newblock (\doi{10.1021/acsmacrolett.5b00511}).

\bibitem{R-Selinger2016}
Konya, A., Gimenez-Pinto, V. \& Selinger, R. L.~B., 2016 Modeling defects,
  shape evolution, and programmed auto-origami in liquid crystal elastomers.
\newblock \emph{Frontiers in Materials} \textbf{3}, 24--30.
\newblock ISSN 2296-8016.
\newblock (\doi{10.3389/fmats.2016.00024}).

\bibitem{modes2016review}
Modes, C. \& Warner, M., 2016 Materials with programmable shapes.
\newblock \emph{Physics Today} \textbf{69}, 32--38.
\newblock (\doi{10.1063/PT.3.3051}).

\bibitem{Plucinsky_2016}
Plucinsky, P., Lemm, M. \& Bhattacharya, K., 2016 Programming complex shapes in
  thin nematic elastomer and glass sheets.
\newblock \emph{Physical Review E} \textbf{94}, 010701(R).
\newblock (\doi{10.1103/PhysRevE.94.010701}).

\bibitem{ware2015voxelated}
Ware, T.~H., McConney, M.~E., Wie, J.~J., Tondiglia, V.~P. \& White, T.~J.,
  2015 Voxelated liquid crystal elastomers.
\newblock \emph{Science} \textbf{347}, 982--984.

\bibitem{sanchez2011valve}
S\'anchez-Ferrer, A., Fischl, T., Stubenrauch, M., Albrecht, A., Wurmus, H.,
  Hoffmann, M. \& Finkelmann, H., 2011 Liquid-crystalline elastomer microvalve
  for microfluidics.
\newblock \emph{Advanced Materials} \textbf{23}, 4526--4530.
\newblock (\doi{10.1002/adma.201102277}).

\bibitem{Bhattacharya2005machine}
Bhattacharya, K. \& James, R.~D., 2005 The material is the machine.
\newblock \emph{Science} \textbf{307}, 53--54.
\newblock (\doi{10.1126/science.1100892}).

\bibitem{warner2003liquid}
Warner, M. \& Terentjev, E.~M., 2003 \emph{Liquid crystal elastomers}, volume
  120.
\newblock Oxford University Press.

\bibitem{van2007glassy}
Van~Oosten, C., Harris, K., Bastiaansen, C.~W. \& Broer, D., 2007 Glassy
  photomechanical liquid-crystal network actuators for microscale devices.
\newblock \emph{The European Physical Journal E: Soft Matter and Biological
  Physics} \textbf{23}, 329--336.
\newblock (\doi{10.1140/epje/i2007-10196-1}).

\bibitem{Mostajeran2017dragging}
Mostajeran, C., Warner, M. \& Modes, C.~D., 2017 Frame{,} metric and geodesic
  evolution in shape-changing nematic shells.
\newblock \emph{Soft Matter} \textbf{13}, 8858--8863.
\newblock (\doi{10.1039/C7SM01596H}).

\bibitem{Priimagi_Iris_2017}
Zeng, H., Wani, O.~M., Wasylczyk, P., Kaczmarek, R. \& Priimagi, A., 2017
  Self-regulating iris based on light-actuated liquid crystal elastomer.
\newblock \emph{Advanced Materials} pp. 1701814--n/a.
\newblock ISSN 1521-4095.
\newblock (\doi{10.1002/adma.201701814}).
\newblock 1701814.

\end{thebibliography}

\end{nolinenumbers}
\end{document}